\documentclass[a4paper,12pt]{article}
\usepackage[T2A]{fontenc}
\usepackage[utf8]{inputenc}

\usepackage[a4paper]{geometry}
\usepackage{amsmath}

\usepackage{graphicx}
\usepackage[english]{babel}
\usepackage{authblk}

\title{\uppercase{The quantum rainbow scattering effect on a single crystalline plane in approximation of continuous potential}}
\author[1,2]{N.F.~Shul'ga$^{*,}$}

\usepackage{lipsum}

\newcommand\blfootnote[1]{%
  \begingroup
  \renewcommand\thefootnote{}\footnote{#1}%
  \addtocounter{footnote}{-1}%
  \endgroup
}

\author[1,2]{V.D.~Koriukina $^{**,}$}
\affil[1]{National Science Center ``Kharkiv Institute of Physics and Technology'', Kharkiv, Ukraine}
\affil[2]{V.N. Karazin Kharkiv National University, Kharkiv, Ukraine}
\begin{document}
\maketitle

\begin{abstract}
{The differential scattering cross section for charged relativistic particles moving parallel close to the crystalline plane of atoms was obtained. The rainbow scattering effect in the approximation of continuous potential was demonstrated. The problem was considered on the basis of the eikonal approximation of quantum electrodynamics.
\par}

\end{abstract}

\section*{{Introduction}}
\blfootnote{$^{*}$shulga@kipt.kharkov.ua}
\blfootnote{$^{**}$koriukina@kipt.kharkov.ua (corresponding author)}

{Starting with M.L. Ter-Mikaelyan's works (see the book \cite{MLT72} and references therein), coherent and interference effects in fast charged particles scattering are of particular interest. Usage of the eikonal and Born approximation of quantum electrodynamics \cite{AIA65} has solved number of scattering problems \cite{AIA96} such as fast charged particles scattering on crystalline strings of atoms, the rainbow scattering effect when particles incedent close to a crystalline string of atoms, and others. Also the idea of continuous potential suggested by J. Lindhard \cite{JLi} allowed to consider complex potentials with less effort.  In the works \cite{Born, Eik}, there was formulated and applied an approach, which allows considering scattering on targets of various configurations from a single point of view. The suggested approach makes it possible to describe scattering on targets with complicated structure in a relatively simple way, creating an alternative to the method of direct simulation of particles passage through the target. The given approach demonstrates the possibility of considering scattering problems both on the basis of the eikonal and Born approximations. The Born approximation makes it relatively simple to obtain the scattering cross section, splitted into coherent and incoherent scattering cross sections. However, the application region of the Born approximation does not allow to use it as the target thickness increases. Starting from a certain thickness of the target (determined by the conditions of application of these approximations), it is necessary to use the eikonal approximation, which has a wider region of application.\par}
{In this work, consideration of fast charged particles scattering on the crystal plane of atoms \cite{Born, Eik}  in the eikonal approximation of quantum electrodynamics was continued. The main attention is paid to obtaining the result in the eikonal approximation without using the stationary phase method, in contrast to the work \cite{Eik}. It was also shown that the obtained result can be approximated near the rainbow point, demonstrating the presence of the rainbow scattering effect for this problem. \par}

\section{{The problem formulation}}
 
{In this work, we investigate the fast charged particles scattering on one crystalline plane of atoms. The plane is chosen so that inside it the particles are distributed with equal probability, thermal fluctuations occur in the transverse to the plane direction. This plane corresponds to a two-dimensional amorphous medium. This problem is a certain approximation for considering the movement of particles in a crystal close to one of its crystalline planes. For high-energy particles, this problem makes sense because the screened potential of atoms leads to a rapid decrease of the total plane potential with distancing from it, and, as a result, the particles scattering is mainly determined with the potential of the nearest plane. Let us consider the simplest version of the crystal orientation to the incident beam: let the particle incident parallel to the plane.

\begin{figure}[ht]
	\centering
			\includegraphics[scale=0.7]{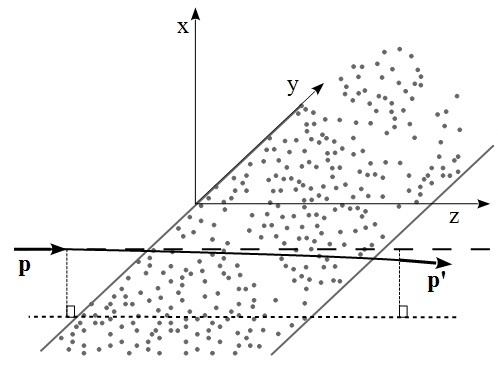}
	\caption{The motion of a fast charged particle incident parallel to the crystal plane}
	\label{FIG:1}
\end{figure}

\par}
{The rainbow scattering on a crystalline string of atoms was considered in \cite{AIA96}. Partly the particles scattering on the specified plane was presented in the works of the authors \cite{Born, Eik}. In the first of these papers \cite{Born}, the problem was a studyed in the Born approximation of quantum electrodynamics. This approximation allows dividing the scattering cross section into coherent and incoherent components. We note that, in contrast to scattering on crystal strings, when scattering on the specified plane, the Debye-Waller factor in the incoherent part is absent in the cross-section. Also, this cross-section differs from the scattering cross-section for a three-dimensional amorphous medium in containing a coherent part. \par}

{Since the Born approximation becomes rapidly inappropriate with target thickness increasing, it makes sense to use the eikonal approximation for the problem as it was done in the second of these works \cite{Eik}. The work \cite{Eik} demonstrated the asymmetry of scattering in the $x$ and $y$ directions. It has been shown that the particles scatter in the $y$ direction similar to how they would scatter in an amorphous medium, while in the $x$ direction the continuous potential of the crystal plane is essential for scattering. \par}
{In this work, we continue to consider the problem of fast charged particles scattering on one crystalline plane with a uniform distribution of particles inside and with thermal deviations of particles positions in the $x$ axis direction. The scattering analysis is based on the approach described in the works  \cite{Born, Eik} in the eikonal approximation of quantum electrodynamics. In contrast to paper \cite{Eik}, we take into account the interference of scattering amplitudes to demonstrate theoretically the quantum rainbow effect for this problem. \par}

\section{{The differential scattering cross section}}
{ \par}
{In the zero eikonal approximation of quantum electrodynamics, the differential scattering cross section is given by the formula \cite{AIA65,AIA96} 
\begin{eqnarray} \label{eq1}
\frac{d^2\sigma}{d\vec{q}_{\perp}}=|a|^2
\end{eqnarray}
where the scattering amplitude $a$ is defined as follows
\begin{eqnarray}  \label{eq2}
a=\frac{i}{2 \pi}\int _{-\infty}^{\infty} d^2\rho \ e^{\frac{i}{\hbar} \vec{q} \vec{\rho}} \left[1- e^{\frac{i}{\hbar} {\chi}_0^{(N)}(\vec{\rho})}  \right]
\end{eqnarray}
where $\vec{\rho}$ are the coordinates transverse to the particles motion, $\vec{q}$ is the momentum transmitted due to scattering, the $\chi_0^{(N)}$ function for $N$ particles in the target is defined  in this way
\begin{eqnarray}  \label{eq3}
\chi_0^{(N)}=-\frac{1}{v} \int _{-\infty}^{\infty} dz \ U^{(N)}(\vec{\rho},z)
\end{eqnarray}
where $v$ is the incident particles velocity, $z$ is the longitudinal coordinate relative to the particles movement, $U^{(N)}$ is the total potential of $N$ particles of the target. \par}
{Since the real potential is very complicated, as an approximation we average the differential scattering cross section over the particles distribution in the target according to the method used in the paper \cite{Eik}. It can be shown that preserving only the terms in the scattering amplitude, which are linear in the function $\chi_0^{(1)}$, we can obtain the following formula for the cross section
 
\begin{eqnarray}  \label{eq4}
a=\frac{-i}{2 \pi}\int _{-\infty}^{\infty} d^2\rho \ e^{\frac{i}{\hbar} \left\{ \vec{q} \vec{\rho} +N \bar{\chi}_0(\vec{\rho}) \right\} } 
\end{eqnarray}

where $\bar{\chi}_0=\bar{\chi}_0^{(1)}$ and the bar corresponds to averaging. \par}
{Here a continuous potential idea \cite{JLi} naturally appears. \par}
{If we assume that the function $\bar{\chi}_0$ depends only on the coordinate $x$ (we will demonstrate the admissibility of this statement later), it is possible, by integrating over $y$, to rewrite the scattering amplitude in this way
\begin{eqnarray}  \label{eq5}
a=\frac{-i}{2 \pi} \int _{-\infty}^{\infty} dx \ e^{\frac{i}{\hbar} \left\{ q_x x +N\bar{\chi}_0\left(x\right) \right\}} \ 2 \pi \delta(q_y)
\end{eqnarray}

Since the square of the Dirac's delta function is
\begin{eqnarray}  \label{eq6}
\left|\int _{-\infty}^{\infty} dy \ e^{\frac{i}{\hbar} q_y y} \right|^2 = 2 \pi L_y \ \delta(q_y)
\end{eqnarray}
where $L_y$ is the size of the target in the $y$ direction, then integrating the differential cross section by the transmitted momentum $q_y$, we come to the this expression for the differential scattering cross section
\begin{eqnarray}  \label{eq7}
\frac{d\sigma}{dq_x}=\frac{L_y}{2 \pi} \ |\tilde{a}|^2
\end{eqnarray}
where 
\begin{eqnarray}  \label{eq8}
\tilde{a}=\int _{-\infty}^{\infty} dx \ e^{\frac{i}{\hbar} \left\{ q_x x +N\bar{\chi}_0\left(x\right) \right\}} 
\end{eqnarray}

Analogically to the rainbow scattering on a string of atoms shown in \cite{AIA96}, we demonstrate the presence of this effect in scattering on a crystal plane. We note that the rainbow scattering effect occurs when the transmitted momentum corresponds to two or more values of impact parameter (from a quasi-classical point of view). In the paper \cite{Eik} we demonstrated that such multiple correspondence exists for this crystalline plane. The rainbow scattering effect is mainly due to  values which are close to the extremum of the dependence function of the transmitted momentum on the impact parameter. Since, from the standpoint of the quasi-classical approach, the transmitted momentum is determined by the derivative $\partial_x\bar{\chi}_0$, then the extremum of this dependence function corresponds to the impact parameter $x_m$ for which $ \partial_x^2\bar{\chi}_0 \left(x_m\right) = $0. Expanding the underexponential expression in \eqref{eq8} near this extremum point $x=x_m+\Delta_x$, we obtain the following series

\begin{eqnarray}  \label{eq9}
q_x x +N\bar{\chi}_0\left(x\right) \approx  q_x x_m +N\bar{\chi}_0 (x_m) + \Delta_x \left[ q_x+N \partial_x\bar{\chi}_0 (x_m) \right]+ N \frac{\Delta_x^3}{6}\partial_x^3\bar{\chi}_0 (x_m)
\end{eqnarray}
Then $\tilde{a}$ is determined by the integral
\begin{eqnarray}  \label{eq10}
& \tilde{a}=e^{\frac{i}{\hbar} \left\{ q_x x_m +N\bar{\chi}_0\left(x_m \right)   \right\}}  \int _{-\infty}^{\infty} d \Delta_x \ exp \left[ \frac{i}{\hbar} \left\{  \Delta_x \left[q_x+N \partial_x\bar{\chi}_0 (x_m) \right]+  N\frac{\Delta_x^3}{6}\partial_x^3\bar{\chi}_0 (x_m)   \right\} \right]
\end{eqnarray}
After integration we obtain the final expression for scattering near the extremum point 
\begin{eqnarray}  \label{eq11}
\tilde{a} =2 \pi   \ e^{\frac{i}{\hbar} \left\{ q_x x_m +N\bar{\chi}_0 (x_m)   \right\}} \left(\frac{N}{2\hbar}\partial_x^3\bar{\chi}_0 (x_m) \right)^{-1/3} \times  \nonumber\\
\times Ai \left[ \frac{q_x+N \partial_x\bar{\chi}_0 (x_m)}{\hbar}  \left(\frac{N}{2\hbar}\partial_x^3\bar{\chi}_0 (x_m) \right)^{-1/3} \right]
\end{eqnarray}

where $Ai$ is the Airy function. So, the differential scattering cross section near the extremum has the following form
\begin{eqnarray}  \label{eq12}
 \frac{d\sigma}{dq_x } =  2 \pi L_y  \ \left(\frac{N}{2\hbar}\partial_x^3\bar{\chi}_0 (x_m) \right)^{-2/3}Ai^2 \left[ \frac{q_x+N \partial_x\bar{\chi}_0 (x_m)}{\hbar}  \left(\frac{N}{2\hbar}\partial_x^3\bar{\chi}_0 (x_m) \right)^{-1/3} \right]
\end{eqnarray}

The scattering cross section far from the extremum point in the quasi-classical approach can be expressed as follows \cite{Eik}  
\begin{eqnarray}  \label{eq13}
& \frac{d\sigma}{dq_x } = L_y \left[ \frac{1}{\left| N \partial_x^2\bar{\chi}_0^{(1)} \right|} + \frac{1}{\left| N \partial_x^2\bar{\chi}_0^{(2)} \right|}\right]  \nonumber\\
& x_i, i=1,2: q_x+ N \partial_x \bar{\chi}_0 (x_i) =0
\end{eqnarray}

where $\bar{\chi}_0^{(i)}=\bar{\chi}_0(x_i)$. \par}

\section{{Functions related to the potential}}
{In order to obtain a specific form of the differential cross section for our case, we need to find the functions contained in the cross section expression. Such functions are listed below. \par}
{The potential of an individual atom in a target with a nuclear charge $Z|e|$ is considered to be a screened Coulomb potential with a screening radius $R$
\begin{eqnarray} 
u(\vec{r})=\frac{Z|e|}{r} \ exp \left[ - \frac{r}{R} \right]
\end{eqnarray}
The $\chi_0$ function for this potential, mentioning that for ultrarelativistic particles $v \rightarrow c$, is
\begin{eqnarray} 
\chi_0=- \frac{ e}{c}\int _{-\infty}^{\infty} dz \: u(r)=\pm 2Z \alpha \: K_0 \left(\frac{\rho}{R} \right)
\end{eqnarray}
where $\alpha$ is a fine structure constant, $K_0$ is a modified Bessel function of the second kind. The sign "$+$" corresponds to the function for electrons, "$-$" for positrons. \par}
{ Let us obtain the averaged $\bar{\chi}_0$ function using its Fourier components $\chi_{\mu}$
\begin{eqnarray} 
\chi_\mu=\frac{\pm 2Z \alpha}{\mu^2+ R^{-2}}
\end{eqnarray}

\begin{eqnarray} 
\bar{\chi}_0=
 \int du_x  \ dy_0 \ f(u_x,y_0) \int_{-\infty}^{\infty} \frac{d^2 \mu}{2 \pi} \:  e^{i {\mu}_x \left(x - {u}_x \right)+i {\mu}_y \left(y - {y}_0 \right)}  \chi_\mu
\end{eqnarray}
where $f$ is the distribution function of atoms in the target, which in the $y$ direction corresponds to the equally probable distribution of atoms and in the $x$ direction is determined by thermal fluctuations:
\begin{eqnarray} 
f(u_x,y_0)=\ \frac{\exp \left[ -\frac{u_x^2}{2 \overline{u_x^2}} \right]}{\sqrt{2\pi \overline{u_x^2}}} \frac{1}{L_y}, \ y_0 \in \left[-\frac{L_y}{2},\frac{L_y}{2} \right]
\end{eqnarray}
After averaging, the $\bar{\chi}_0$ function is
\begin{eqnarray} 
\bar{\chi}_0= \pm Z \alpha \ \frac{\pi R}{L_y} \ e^{\frac{\overline{u_x^2}}{2R^2}} \left\{ e^{\frac{x}{R}} \: Erfc \left[ \sqrt{\frac{\overline{u_x^2}}{2R^2}} +\frac{x}{\sqrt{2\overline{u_x^2}}} \right]  + e^{-\frac{x}{R}} \: Erfc \left[ \sqrt{\frac{\overline{u_x^2}}{2R^2}} -\frac{x}{\sqrt{2\overline{u_x^2}}} \right]    \right\}
\end{eqnarray}
The fact that the $\bar{\chi}_0$ function does not depend on $y$ is in good agreement with the equally probable distribution of atoms in this direction. Derivatives of the $\bar{\chi}_0$ function are also required to obtain the scattering cross section. The first derivative from a quasi-classical point of view determines the transmitted momentum $\left( q_x=-N \ \partial_x \bar{\chi}_0 \right)$ and has the following form:
\begin{eqnarray} 
\partial_x\bar{\chi}_0=  \frac{ \pm \pi Z \alpha }{L_y} \ e^{\frac{\overline{u_x^2}}{2R^2}}  \left\{ e^{\frac{x}{R}} \: Erfc \left[ \sqrt{\frac{\overline{u_x^2}}{2R^2}} +\frac{x}{\sqrt{2\overline{u_x^2}}} \right] - e^{-\frac{x}{R}} \: Erfc \left[ \sqrt{\frac{\overline{u_x^2}}{2R^2}} -\frac{x}{\sqrt{2\overline{u_x^2}}} \right]    \right\}
\end{eqnarray}

The second and third derivatives have the following expressions, respectively
\begin{eqnarray} 
\partial^2_x\bar{\chi}_0= \frac{\pm \pi Z \alpha}{RL_y} \ e^{\frac{\overline{u_x^2}}{2R^2}} \left\{ e^{\frac{x}{R}} \: Erfc \left[ \sqrt{\frac{\overline{u_x^2}}{2R^2}}+\frac{x}{\sqrt{2\overline{u_x^2}}} \right]  + \right.  \nonumber\\ 
\left. + e^{-\frac{x}{R}} \: Erfc \left[ \sqrt{\frac{\overline{u_x^2}}{2R^2}} -\frac{x}{\sqrt{2\overline{u_x^2}}} \right]  -\frac{2\sqrt{2}\ R}{\sqrt{\pi \overline{u_x^2}}} \ exp \left[-\frac{\overline{u_x^2}}{2 R^2}-\frac{x^2}{2 \overline{u_x^2}} \right] \right\}
\end{eqnarray}

\begin{eqnarray} 
\partial^3_x\bar{\chi}_0= \frac{\pm \pi Z \alpha}{R^2L_y} \ e^{\frac{\overline{u_x^2}}{2R^2}} \left\{e^{\frac{x}{R}} \: Erfc \left[ \sqrt{\frac{\overline{u_x^2}}{2R^2}} +\frac{x}{\sqrt{2\overline{u_x^2}}} \right] - \right.  \nonumber\\ 
\left.  -e^{-\frac{x}{R}} \: Erfc \left[ \sqrt{\frac{\overline{u_x^2}}{2R^2}} -\frac{x}{\sqrt{2\overline{u_x^2}}} \right]  +\frac{2\sqrt{2}\ R^2 x}{\overline{u_x^2} \sqrt{\pi \overline{u_x^2}}} \ exp \left[-\frac{\overline{u_x^2}}{2 R^2}-\frac{x^2}{2 \overline{u_x^2}} \right] \right\}
\end{eqnarray}
\par}


\section{{Calculation results for the differential scattering cross section. Rainbow scattering effect}}
{Using the $\bar{\chi}_0$ function and its derivatives, we obtain the differential scattering cross sections for the problem. Let the target consist of silicon atoms. \par}

\begin{figure}[!ht]
	\centering
		\includegraphics[scale=0.7]{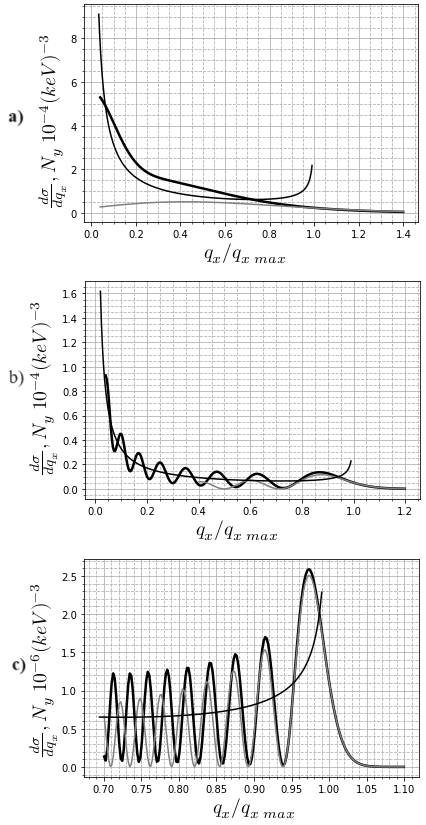}	
	\caption{The differential cross section of the fast charged particles scattering on the crystal plane of atoms for different $nL_z$ parameters: the bold line corresponds to the numerical integration according to the formulas \eqref{eq7}, \eqref{eq8}, thin black line to \eqref{eq13}, thin gray line to \eqref{eq12}; figure \textbf {a)} is for $nRL_z=10$, \textbf {b)} for $nRL_z=100$, \textbf {c)} for $nRL_z=1\ 000$  }
	\label{FIG:2}
\end{figure}

{Numerically integrating the formula \eqref{eq8} and substituting it into \eqref{eq7}, we obtain the differential scattering cross section. During the numerical integration, the infinite limits of the integral \eqref{eq8} are replaced by finite ones, which causes an error in the determination of small-angle scattering. However, this error does not make a dominant contribution since the target potential is short-ranged and decreases rapidly. The obtained result is similar to the result of particle scattering on a crystalline string of atoms \cite{AIA96}. \par}
{This result can be approximated by the formula \eqref{eq12} near the extremum of the transmitted momentum dependence on impact parameter and by the formula \eqref{eq13} with distancing from this point. Since the second derivative becomes zero at the point of extremum of the first derivative (and as a result, at the maximum classically allowed momentum $q_{x \ max}$), according to \eqref{eq13} the differential cross section increases infinitely near $q_x=q_{x \ max} $, which is not physical. The formula \eqref{eq12} "removes" \ the infinite growth of the cross section, and also well approximates the result of numerical integration beyond the limits of classically allowed transmitted momenta. \par}
{We note that the expressions of the form $N \partial_x^n \bar{\chi}_0, \ n=0,1,2,3 ...$ are proportional to $N/L_y$ and do not contain the values $N, \ L_y$ separately. Therefore, with accuracy up to the constant multiplier $L_y$, to which the section is proportional, the features of the cross section are determined by the value $N/L_y$. This value could be also expressed in terms of longitudinal size of the plane $L_z$ with the help of the ratio 
\begin{eqnarray} 
N=n L_y L_z
\end{eqnarray}
where n is the density of atoms in the plane. So, we may use $nL_z$ as the parameter instead of $N/L_y$ to show how the scattering features depend on the target thickness.  Figure \ref{FIG:2} presents calculations for different values of the parameter $nL_z$. $N_y$ on the figure is defined as $N_y=L_y/a$, where $a$ is average distance between atoms in target. \par}
{Also, it is important to take into account the conditions of application of the eikonal approximation for the specified problem, which are as follows:

\begin{eqnarray} 
U^{(N)}<<pv, \ nL_z \tilde{\chi}_0 << p, \  \left(nRL_z \right)^2 \frac{\tilde{\chi}_0}{pR}<<1
\end{eqnarray}
where 
\begin{eqnarray} 
\tilde{\chi}_0 =  \pm Z \alpha \pi  \ e^{\frac{\overline{u_x^2}}{2R^2}}  \left\{ e^{\frac{x}{R}} \: Erfc \left[ \sqrt{\frac{\overline{u_x^2}}{2R^2}} +\frac{x}{\sqrt{2\overline{u_x^2}}} \right]  + e^{-\frac{x}{R}} \: Erfc \left[ \sqrt{\frac{\overline{u_x^2}}{2R^2}} -\frac{x}{\sqrt{2\overline{u_x^2}}} \right]    \right\}
\end{eqnarray}
\par}

{ \par}
{ \par}

\section*{Conclusions}

{In the present paper, we considered the problem of fast charged particles scattering on the crystal plane of atoms in the eikonal approximation, which allows us to excede the limits of the Born approximation. Extending our previous research \cite{Born, Eik}, we obtained the differential scattering cross section for the specified problem, avoiding the usage of an additional approximation (stationary phase method). Taking into account the configuration of the crystal plane made it possible to reduce the scattering problem to a one-dimensional one. It was also shown that near the extremum of the transmitted momentum dependence on the impact parameter, the scattering cross section has an approximation that indicates the presence of the rainbow scattering effect. We note that omitting $\chi_0^2$ terms (which took place in \eqref{eq4}) leads us to approximation of continuous potential which does not count the incoherent scattering. The incoherent scattering in its turn is of particular interest since it may impact the scattering cross section.  \par}
{Also, the differential scattering cross sections for various target parameters were obtained. These results demonstrate how the frequency of cross section oscillations near the rainbow point is related to the target thickness. The application regions of the eikonal approximation for the given problem are also specified. \par}

\section*{Acknowledgements}
{The work was partially supported by the National Academy of \ Sciences of Ukraine (project 0121U111556) and by Agence Universitaire de la Francophonie.
\par}

\end{document}